\documentclass[11pt]{article}
\usepackage[margin=1in]{geometry}
\usepackage[T1]{fontenc}
\usepackage{lmodern}
\usepackage{microtype}
\usepackage{amsmath}
\usepackage{booktabs}
\usepackage{graphicx}
\usepackage{xcolor}
\usepackage{tikz}
\usetikzlibrary{positioning,arrows.meta,fit}
\usepackage{xurl}
\usepackage[colorlinks=true,linkcolor=black,citecolor=black,urlcolor={blue!65!black}]{hyperref}
\usepackage[numbers,sort&compress]{natbib}

\title{Paradee: Distilling Kokoro-82M into an\\8M-Parameter Single-Voice Text-to-Speech Model}
\author{Sahil Mahendrakar}
\date{September 2026}

\begin{document}
\maketitle

\begin{abstract}
We distill Kokoro-82M, a widely used open text-to-speech model with 54 voices, into Paradee, an 8.07M-parameter model that speaks one of them.
Paradee keeps Kokoro's architecture with much narrower layers, and each of its two halves is trained separately against the frozen teacher.
It has 10$\times$ fewer parameters and needs 15$\times$ less compute.
We first synthesize a corpus with the teacher and keep its durations, pitch, energy and phoneme features.
We then train a small text side to predict these values, and a small decoder to turn the teacher's saved values into the teacher's audio, first with spectral losses and then adversarially.
Finally, we connect the two halves and quantize the weights to int8.
It needs no alignment learning and no joint training, and it runs on one laptop.
Stored in int8, Paradee is 8.5\,MB, runs 25$\times$ faster than real time on one CPU thread, and scores 4.41 on UTMOS against the teacher's 4.52.
The student initially kept a slight buzz, which we trace to the phase of voiced speech between 2 and 8\,kHz.
A phase-locking filter applied after synthesis removes most of it, with no training and no extra parameters.
Code, model files and audio samples are at \href{https://github.com/sahilmahendrakar/paradee}{github.com/sahilmahendrakar/paradee}.
\end{abstract}

\section{Introduction}

Text-to-speech that runs on the device, without a server, has to be small enough to ship, fast enough to run faster than real time on a CPU, and natural enough to listen to for long stretches.
Kokoro-82M~\cite{hexgrad2025kokoro} is a widely used open-weight text-to-speech model with high naturalness for its size.
It is derived from StyleTTS~2~\cite{li2023styletts2} and ships 54 voices, but it has 82 million parameters and needs 55 GFLOP per second of audio.

Many applications only ever use one voice.
We ask how small Kokoro can get if it has to speak just one voice, and whether that can be done cheaply.
Its authors trained it with about 1{,}000 A100 GPU-hours~\cite{hexgrad2025kokoro}, whereas we want a student that can be distilled and run on one laptop.

Our approach follows the structure of the teacher (Figure~\ref{fig:pipeline}).
Kokoro has two halves.
The text side predicts durations, pitch, energy and phoneme features, and the decoder turns them into audio.
We first run the teacher on 12{,}000 English sentences.
For each one we save the audio it produced along with the values it computed on the way, namely the duration of each phoneme, the pitch and energy contours, and the phoneme features passed to the decoder.
These saved outputs are the training data for both halves of the student.

We then shrink each half and train it separately against the teacher, which stays frozen, meaning its weights are never updated and it only supplies targets.
The student text side reads the phonemes and learns to predict the teacher's saved durations, pitch, energy and phoneme features.
The student decoder is given the teacher's saved durations, pitch, energy and phoneme features as input and learns to reproduce the teacher's saved audio from them.
Because each half trains against fixed teacher outputs, no alignment has to be learned.

To join them, we connect the student text side's outputs to the student decoder's inputs.
At inference the text side reads the phonemes and predicts a duration for each one.
These durations stretch its phoneme features to one per audio frame, and it then predicts pitch and energy for every frame.
The decoder turns these into audio, and nothing is retrained after the halves are joined.
Last, we quantize the weights to 8-bit integers, which makes the model four times smaller with no measurable change in UTMOS.

The result, Paradee, has 8.07M parameters (10$\times$ fewer than the teacher's 81.8M) and needs 15$\times$ less compute.
Stored in int8, it is 8.5\,MB, runs 25$\times$ faster than real time on one CPU thread, and scores 4.41 on UTMOS~\cite{saeki2022utmos} against the teacher's 4.52.

Our contributions are:
\begin{enumerate}
  \item \textbf{A simple single-voice distillation method} (Section~\ref{sec:method}). Each half is distilled separately against the frozen teacher and the halves are joined without joint training. It needs no alignment learning and no teacher at inference time, and it trains on one laptop.
  \item \textbf{What we learn from distilling the halves separately} (Section~\ref{sec:results}). Pairing each student half with the other half of the teacher shows what each one costs. A 4.2M text side scores UTMOS 4.36 through the teacher's decoder, and a 7.1M one does no better. The 3.9M decoder scores 4.37 from the teacher's text side. Together the two halves score 4.39, so their errors do not add up. For the decoder, the balance of the training losses matters more than its size. Lowering the weight of the spectral loss so that the adversarial critics could take effect raised UTMOS from 3.0 to 4.4, while making the decoder wider or twice as large did not help. Finally, the text side must be trained to match the teacher's durations, pitch, energy and phoneme features directly. A text side trained instead to make the teacher's decoder sound good reached the teacher's own score with that decoder, but fell to 3.78 with the student decoder, against 4.39 for the directly trained one.
  \item \textbf{A map of where Kokoro spends its parameters and compute} (Section~\ref{sec:anatomy}). The waveform generator is a quarter of the parameters but 89\% of the compute, and the small text side is the part most sensitive to quantization.
  \item \textbf{A diagnosis and fix for the student's residual buzz} (Section~\ref{sec:buzz}). Listening tests place it in the phase of voiced speech between 2 and 8\,kHz, just above where Kokoro's harmonic excitation stops, and a training-free phase-locking filter removes most of it.
\end{enumerate}

\section{Background}
\label{sec:background}

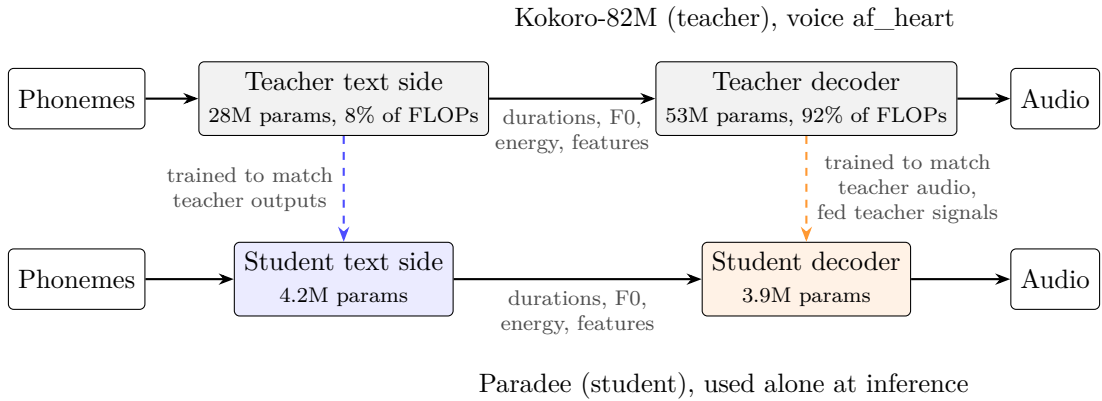
\begin{figure}[t]
\centering
\begin{tikzpicture}[
  box/.style={draw, rounded corners=2pt, minimum height=0.8cm, align=center, font=\small, fill=white},
  arr/.style={-{Stealth[length=2mm]}, thick},
  lab/.style={font=\scriptsize, align=center, text=black!70}]
  \node[box] (ph) {Phonemes};
  \node[box, right=0.7cm of ph, fill=black!5] (tt) {Teacher text side\\\scriptsize 28M params, 8\% of FLOPs};
  \node[box, right=2.2cm of tt, fill=black!5] (td) {Teacher decoder\\\scriptsize 53M params, 92\% of FLOPs};
  \node[box, right=0.7cm of td] (wav) {Audio};
  \draw[arr] (ph) -- (tt);
  \draw[arr] (tt) -- node[lab, below] {durations, F0,\\energy, features} (td);
  \draw[arr] (td) -- (wav);
  \node[font=\small, above=0.25cm of tt.north east, anchor=south west, xshift=0.2cm] {Kokoro-82M (teacher), voice af\_heart};
  \node[box, below=1.4cm of tt, fill=blue!8] (st) {Student text side\\\scriptsize 4.2M params};
  \node[box, below=1.4cm of td, fill=orange!10] (sd) {Student decoder\\\scriptsize 3.9M params};
  \draw[arr, dashed, blue!70] (tt) -- node[lab, left] {trained to match\\teacher outputs} (st);
  \draw[arr, dashed, orange!80] (td) -- node[lab, right] {trained to match\\teacher audio,\\fed teacher signals} (sd);
  \node[box] (sph) at (ph |- st) {Phonemes};
  \node[box] (swav) at (wav |- sd) {Audio};
  \draw[arr] (sph) -- (st);
  \draw[arr] (st) -- node[lab, below] {durations, F0,\\energy, features} (sd);
  \draw[arr] (sd) -- (swav);
  \node[font=\small, below=0.6cm of st.south east, anchor=north west, xshift=0.2cm] {Paradee (student), used alone at inference};
\end{tikzpicture}
\caption{Kokoro's two halves (top) and Paradee's (bottom). Each student half learns from the frozen teacher separately (dashed arrows). At inference the two student halves are connected and turn phonemes into audio on their own, without the teacher.}
\label{fig:pipeline}
\end{figure}

Kokoro follows the StyleTTS~2 design~\cite{li2023styletts2,li2022styletts}.
Text is converted to phonemes, and the model then works in two halves (Figure~\ref{fig:pipeline}, top), both of which receive a style vector that selects the voice.

The \textbf{text side} decides what to say and how.
A phoneme-level BERT~\cite{li2023plbert} with the ALBERT architecture~\cite{lan2020albert} reads the phonemes, and a prosody predictor turns them into a duration for each phoneme, a pitch contour (F0) and an energy contour.
A separate text encoder produces 512-channel features that describe the sound of each phoneme (we call them the \emph{phoneme features}).
After they are stretched by the durations, they are passed to the decoder.

The \textbf{acoustic side}, or decoder, decides what the waveform looks like.
It works in two stages.
First, a stack of residual convolutional blocks with adaptive instance normalization (AdaIN)~\cite{huang2017adain} combines the phoneme features with the pitch and energy contours, conditioned on the style vector.
These blocks account for 41\% of Kokoro's parameters.
Then a waveform generator, an iSTFTNet~\cite{kaneko2022istftnet} built on HiFi-GAN~\cite{kong2020hifigan}, turns the result into audio at 24\,kHz.
To help it produce voiced sounds, the generator is also given a tone computed from the predicted pitch, made of sine waves at the pitch and its first eight overtones~\cite{wang2019nsf}.

\section{Analysis of the teacher}
\label{sec:anatomy}

Before training anything, we measured how Kokoro's parameters and compute are divided among its parts and how sensitive each part is to quantization.
These decide which part must shrink most.
All measurements use Kokoro-82M v1.0 with voice \texttt{af\_heart} on an Apple M4 Pro.

\begin{figure}[t]
\centering
\includegraphics[width=0.6\linewidth]{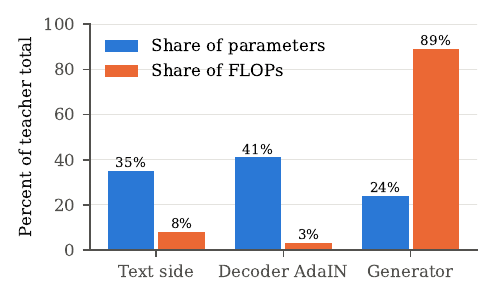}
\caption{Share of Kokoro's parameters and of its compute (FLOPs per second of audio) in each part. The text side is ALBERT, the text encoder and the prosody predictor. The waveform generator is a quarter of the parameters but almost all of the arithmetic.}
\label{fig:anatomy}
\end{figure}

\paragraph{Parameter and compute breakdown.}
Figure~\ref{fig:anatomy} and Table~\ref{tab:census} split the model into three parts.
The text side holds 35\% of the parameters, the decoder's AdaIN blocks 41\% and the generator 24\%.
The compute is split very differently.
The generator performs 89\% of the 55 GFLOP needed per second of audio, because its last stages run convolutions at 60$\times$ the frame rate.
The text side is under 10\%.
So on a CPU, run time is dominated by the generator's per-sample convolutions, not by the parameter count.
Any large speed-up therefore has to come from shrinking the generator, whatever happens to the text side.

\begin{table}[t]
\centering
\small
\begin{tabular}{lrr}
\toprule
Subsystem & Parameters & Share \\
\midrule
Decoder AdaIN encode/decode blocks (1024 channels) & 33.6M & 41\% \\
Decoder iSTFTNet generator & 19.7M & 24\% \\
LSTMs (text encoder and prosody predictor) & 11.2M & 14\% \\
ALBERT (12 shared layers, 768 hidden) and its output projection & 6.7M & 8\% \\
F0, energy and duration heads & 6.6M & 8\% \\
Text encoder CNN and phoneme embedding & 4.0M & 5\% \\
\midrule
Total (PyTorch) & 81.8M & \\
\bottomrule
\end{tabular}
\caption{Parameter census of Kokoro-82M v1.0.}
\label{tab:census}
\end{table}

\paragraph{Quantization sensitivity.}
Storing Kokoro's weights in 8 bits barely changes its output, and every quantized version of the teacher we tried still scored 4.52--4.54 on UTMOS.
The small text side is the most sensitive part, because a slightly changed duration shifts every sound that follows.
The common int8 download of Kokoro sounds worse mainly because it also quantizes activations.
Appendix~\ref{app:quant} gives the full breakdown.

\section{Method}
\label{sec:method}

\subsection{Single-voice specialization}
Paradee is trained for a single voice, \texttt{af\_heart}.
Because the voice is fixed, we remove the style input from both halves of the student and replace it with a learned constant vector.
In Kokoro, the style vector for a given voice also changes with the length of the utterance, so the student has to learn this dependence on length from the training data.

\subsection{Teacher corpus}
We synthesized 12{,}000 English sentences from WikiText-103~\cite{merity2017wikitext} with the teacher, 23.9 hours of audio.
For every sentence we stored the phonemes, the predicted durations, F0 and energy contours, the phoneme features and the waveform (6.9\,GB).
Waveforms were generated on CPU with a fixed seed per sentence so the excitation can be reproduced exactly, since Kokoro's output differs slightly between CPU and GPU.
The first 200 sentences are held out for evaluation.

\subsection{Student architecture}
Each student half is Kokoro's own code with smaller widths, so every student layer has a teacher counterpart to learn from.

\paragraph{Text side (4.23M parameters).}
It has an ALBERT with 256 hidden units, 6 layers and 4 heads, and a prosody predictor and text encoder with 192 channels and 3 layers.
A 2-layer MLP projects the 192-channel text features up to the 512 channels the decoder expects.
We also trained a variant with a linear projection (4.0M) and a wider one (7.1M).

\paragraph{Decoder (3.85M parameters).}
The decoder keeps Kokoro's layout, with 256 channels in the AdaIN blocks instead of 1{,}024 and 128 in the generator instead of 512.
It also keeps Kokoro's upsampling factors, kernel sizes (3, 7, 11), excitation module and iSTFT head.
The whole decoder needs 3.36 GFLOP per second of audio, 15$\times$ less than the teacher's decoder, and its generator has 1.19M parameters.
Table~\ref{tab:decoders} in the appendix lists the other sizes we built.

\subsection{Training}
\paragraph{Text side.} The student reads phonemes and is trained, with the teacher's durations given (teacher forcing), to predict the teacher's log durations, pitch, energy and phoneme features with L1 losses.
We train it for 8{,}000 steps at batch size 32 with AdamW and a learning rate of $5\times10^{-4}$.

\paragraph{Decoder.} The student decoder is fed the teacher's own phoneme features, pitch and energy and trained to reproduce the teacher's waveform on 1.6\,s segments.
Stage one uses only spectral losses, which compare how loud each frequency is over time in the student's audio and the teacher's, ignoring phase.
The first is a log-mel L1 loss, which compares spectrograms on the mel scale, a frequency scale spaced the way human hearing works.
The second is a multi-resolution STFT loss~\cite{yamamoto2020pwg}, which compares ordinary spectrograms at three resolutions (FFT sizes 512, 1024 and 2048).
Stage one runs for 50{,}000 steps at batch size 16 and a learning rate of $2\times10^{-4}$.
Stage two adds HiFi-GAN-style adversarial training~\cite{kong2020hifigan}, in which small discriminator networks learn to tell the student's audio from the teacher's and the student learns to fool them.
We use a multi-period discriminator (periods 2, 3, 5, 7 and 11) and a multi-resolution spectrogram discriminator~\cite{jang2021univnet}, 3.2M parameters together, with least-squares GAN losses and feature matching, at batch size 8.
We ran 5{,}000 steps with the spectral losses weighted 10 times the adversarial loss, then 5{,}000 more with them weighted 3 times.

\paragraph{Assembly.} The two halves are then connected with no joint training, as described in Section~1.

\subsection{Phase-locking filter}
After synthesis, a filter with no parameters corrects the phase of voiced speech between 2 and 8\,kHz.
Section~\ref{sec:buzzfix} describes and evaluates it.

\subsection{Post-training quantization}
Every weight matrix is quantized to int8 per output channel with an fp16 scale, and all other parameters are kept in fp16.
Saved this way, the whole model is an 8.45\,MB file, against 32.5\,MB in fp32.

\section{Experimental setup}

\paragraph{Metrics.}
\emph{DTW log-mel L1} compares the spectrogram of a rendition with the teacher's after aligning the two in time with dynamic time warping (DTW), which locally stretches or compresses one of them so that matching sounds line up.
A rendition that is slightly faster or slower is therefore not penalized for timing alone.
It is good for teacher-forced comparisons where timing is fixed, but it saturates for free-running students, whose valid prosody differs from the teacher's (a 3\% speed change alone scores 0.36).
\emph{UTMOS}~\cite{saeki2022utmos} is a neural network trained on human ratings to predict the mean opinion score (MOS) that listeners would give a clip for naturalness, on a scale from 1 (bad) to 5 (excellent).
We use its \texttt{utmos22\_strong} checkpoint, and it is our main metric for comparing free-running students.
All UTMOS scores are averages over the 200 held-out sentences.
The 95\% confidence interval of the mean is $\pm0.003$ for the teacher and between $\pm0.01$ and $\pm0.04$ for the students.
\emph{Word error rate} (WER) comes from transcribing the audio with Whisper (base) and comparing it with the source text.
It measures intelligibility, and it barely separates our models, so we report it only for the main result.
\emph{Speaker similarity} is the cosine between Resemblyzer speaker embeddings~\cite{resemblyzer} of a system's utterance and the teacher's utterance of the same sentence.
We also compared samples by listening throughout, which the author did informally.

\paragraph{Hardware.}
All training ran on one MacBook Pro (Apple M4 Pro, 24\,GB) using the MPS backend.
Speed is measured on the same machine on one CPU thread, over 8 held-out sentences (about 61 seconds of audio), excluding grapheme-to-phoneme conversion for both models.

\section{Results}
\label{sec:results}

\subsection{Main result}

\begin{table}[t]
\centering
\small
\begin{tabular}{lrr}
\toprule
 & Kokoro-82M (teacher) & Paradee (student) \\
\midrule
Parameters & 81.8M & 8.07M (10.1$\times$ fewer) \\
Weights on disk & 325\,MB fp32 ONNX; 82\,MB int8 & 8.45\,MB int8 \\
Compute per second of audio & 55 GFLOP & 3.4 GFLOP (15$\times$ less) \\
Speed, PyTorch, 1 CPU thread & 7.6$\times$ real time & 25.0$\times$ real time \\
Decoder alone, onnxruntime, 1 CPU thread & & 20.5$\times$ real time \\
UTMOS, fp32 & 4.52 $\pm$ 0.003 & 4.41 $\pm$ 0.01 \\
UTMOS, int8 & & 4.41 $\pm$ 0.01 \\
Whisper WER & 5.7\% & 5.7\% \\
Speaker similarity to the teacher & & 0.958 $\pm$ 0.002 \\
\bottomrule
\end{tabular}
\caption{Paradee against its teacher. The student uses the 4.23M text side (feature-supervised, MLP projection), the 3.85M decoder after both training stages, and the phase-locking filter. UTMOS, WER and speaker similarity are over 200 held-out sentences, with 95\% confidence intervals.}
\label{tab:main}
\end{table}

Table~\ref{tab:main} compares Paradee with its teacher.
With a tenth of the parameters and a fifteenth of the compute, Paradee scores 4.41 on UTMOS against the teacher's 4.52, has the same word error rate, and runs 3.3 times faster on one CPU thread.
Speed improves less than compute (3.3$\times$ against 15$\times$) because at this size PyTorch's per-operation overhead on small convolutions and LSTMs dominates.
The phase-locking filter takes 5\% of the run time.
Int8 weights cost nothing measurable (UTMOS 4.41 in both cases), and intelligibility is the same as the teacher's (WER 5.7\% for both).
Quantizing to 4 bits, however, lowered the decoder's UTMOS to 3.98 in an earlier test on 3 sentences, although the teacher's much larger decoder blocks changed little at 4 bits (Appendix~\ref{app:quant}).
A small model has no redundant channels left to absorb the rounding.

Paradee keeps the teacher's voice.
Its speaker similarity to the teacher on the same sentences is 0.958.
For reference, a different Kokoro voice, af\_bella, reading the same sentences scores 0.842.

\begin{table}[t]
\centering
\small
\begin{tabular}{lrrrrr}
\toprule
Model (voice) & Params & File size & Speed & UTMOS & WER \\
\midrule
Kokoro-82M, teacher (af\_heart) & 81.8M & 325\,MB & 7.6$\times$ & 4.52 $\pm$ 0.003 & 5.7\% \\
Paradee (af\_heart) & 8.07M & 8.45\,MB & 25.0$\times$ & 4.41 $\pm$ 0.01 & 5.7\% \\
Paradee, released ONNX file & 8.07M & 9.0\,MB & 17.8$\times$ & 4.41 & 6.0\% \\
Kokoro-7M-Distill (af\_msa) & 7.48M & 30.1\,MB & 35.5$\times$ & 4.18 $\pm$ 0.03 & 7.4\% \\
Piper, en\_US-lessac-medium & 15.7M & 63.2\,MB & 15.4$\times$ & 4.36 $\pm$ 0.02 & 8.8\% \\
KittenTTS nano 0.8 (Bella) & 14.0M & 56.8\,MB & 10.7$\times$ & 4.01 $\pm$ 0.04 & 5.5\% \\
\bottomrule
\end{tabular}
\caption{Paradee against other small open models on the same 200 held-out sentences, with the same scoring. File size is the model as distributed (Paradee int8, the others fp32). Speed is real-time factor on one CPU thread of the same machine. Paradee and the Kokoro models are timed from phonemes. Piper and KittenTTS are timed from text, so their times include phonemization. Each model speaks its own voice, so UTMOS compares naturalness, not closeness to af\_heart.}
\label{tab:baselines}
\end{table}

Table~\ref{tab:baselines} compares Paradee with three small open models.
It also lists the released version of Paradee, a single int8 ONNX file with the phase-locking filter built in, measured on the same sentences.
It is slightly larger and slower than our PyTorch model because of how the model and filter are packaged, and its quality is the same.
Paradee scores highest on UTMOS among them and has the smallest file.
Kokoro-7M-Distill~\cite{oddadmix2026kokoro7m}, the closest in design, runs about 1.4 times as fast but scores 0.23 lower on UTMOS, and KittenTTS~\cite{kittenml2025kittentts} has the lowest word error rate.

\subsection{Text-side students}

\begin{table}[t]
\centering
\small
\begin{tabular}{lrrr}
\toprule
Text side, rendered by the teacher decoder & Params & DTW log-mel & UTMOS \\
\midrule
Teacher text side & 28M & 0.12 (floor) & 4.52 \\
Student s, linear projection & 4.0M & 0.97 & 4.15 \\
Student s, MLP projection & 4.2M & 0.91 & 4.36 \\
Student m, MLP projection & 7.1M & 0.89 & 4.35 \\
Student s, MLP, trained through the teacher decoder & 4.2M & 0.99 & 4.52 \\
\bottomrule
\end{tabular}
\caption{Text-side students, each rendered by the frozen teacher decoder on the 200 held-out sentences. The teacher row's DTW is the teacher against a second run of itself with a different random seed. The last row adds a log-mel loss on audio produced by passing the student's outputs through the frozen teacher decoder. It has the worst DTW but the best UTMOS, because it learns prosody that sounds natural but differs from the teacher's rendition.}
\label{tab:text}
\end{table}

We first tested each student text side by plugging it into the frozen teacher decoder (Table~\ref{tab:text}).

\paragraph{Effect of size.}
Growing the text side from 4.2M to 7.1M improved every teacher-forced loss slightly but left the free-running result unchanged (DTW 0.91 against 0.89, UTMOS 4.36 against 4.35).
The remaining gap is set by the training objective, not capacity.

\paragraph{Feature projection.}
The teacher's 512 phoneme-feature channels are highly redundant.
About 190 dimensions hold 90\% of their variance and 430 hold 99\%.
A student with 192 channels and a linear projection can express at most 192 of those dimensions, and it came close to the best error any such projection allows (0.079 against 0.068), so the projection, not the student, was the limit.
Replacing the linear projection with a small two-layer network (an MLP) removed this limit.
Its error fell to 0.066, below what any linear projection can reach, and UTMOS rose from 4.15 to 4.36.
To see where the remaining error comes from, we took one kind of output at a time from the student and the rest from the teacher, and measured the distance to the teacher's audio on 8 held-out sentences.
Compared with the teacher's own floor of 0.12, student phoneme features add about 0.47, student pitch and energy about 0.35, and student durations only about 0.13.

\paragraph{Training through the teacher decoder.}
We also tried giving the text side one more training objective.
Its predictions are passed through the frozen teacher decoder, and the resulting audio is compared with the teacher's recording using a log-mel loss.
This lifted the 4.2M text side to the teacher's own score (4.52).
But those features were specialized to the teacher's decoder.
Connected to Paradee's student decoder, they scored 3.78, while the text side trained on the teacher's values directly scored 4.39 with the same decoder.
Features learned by matching the teacher's features transfer to a new decoder, and features learned by pleasing one particular decoder do not.
Paradee therefore uses the feature-supervised text side.

\subsection{Decoder students}

\begin{table}[t]
\centering
\small
\begin{tabular}{lrr}
\toprule
Decoder A (3.85M) training & Decoder alone & Full student \\
\midrule
Spectral losses only, 50k steps & 2.98 & 3.00 \\
+ adversarial, spectral weight 45, 3k steps & 3.02 & \\
+ adversarial, spectral weight 10, 5k steps & 4.29 & 4.32 \\
+ adversarial, spectral weight 3, 5k more steps (stage two) & 4.37 & 4.39 \\
+ phase-locking filter (Paradee) & & 4.41 \\
\bottomrule
\end{tabular}
\caption{UTMOS of the student decoder (driven by the teacher's text side) and of the full student, over 200 held-out sentences. The teacher scores 4.52.}
\label{tab:decoder}
\end{table}

With spectral losses alone the small decoder converged to a DTW distance of 0.50 but scored only 2.98 on UTMOS (Table~\ref{tab:decoder}).
At this point the full student sounds like the af\_heart voice with a second, robotic voice speaking at the same time.
Spectrograms show why.
The student keeps crisp harmonic stripes up to 6--8\,kHz, whereas the teacher's harmonics dissolve into breathy noise above 3--4\,kHz, so the sine excitation leaks through almost unchanged.
Level, frequency balance and silence were all correct.

Adversarial training fixes most of this, but only once the critics are allowed to matter.
With HiFi-GAN's default spectral weight of 45, the spectral term outweighed the adversarial terms about 15 to 1, and UTMOS barely moved in 3{,}000 steps (2.98 to 3.02).
Lowering the weight to 10 and then 3 raised the decoder to 4.37 and the full student to 4.39.
The phase-locking filter adds little to UTMOS (4.39 to 4.41) but is what removes most of the audible buzz (Section~\ref{sec:buzz}).

\paragraph{Joining the halves.}
With spectral-only decoders, the student text side through the teacher decoder scored 0.87 on DTW, the teacher text side through the student decoder 0.50, and the full student 0.81, better than the text side alone.
The teacher decoder, trained on exact features, amplifies small feature errors.
The student decoder, trained on audio, is more forgiving.
Measuring each half against the other teacher half therefore overstates what the combination loses, and we did not need joint training.
The final models show the same on UTMOS.
The student text side through the teacher decoder scores 4.36 and the teacher text side through the student decoder 4.37, each about 0.15 below the teacher, yet the full student scores 4.39 before the filter.

\section{Residual buzz and the phase-locking filter}
\label{sec:buzz}\label{sec:buzzfix}

After adversarial training, the student's one audible flaw was a slight buzz on voiced sounds.
To locate it, we combined the magnitude of the student's spectrogram with the teacher's phase, and the reverse, using identical excitation.
The student's magnitude with the teacher's phase had no audible buzz (UTMOS 4.49, against 4.37 for the student and 4.52 for the teacher), while the teacher's magnitude with the student's phase still buzzed (4.45).
Swapping in the teacher's phase only between 2 and 8\,kHz was enough to remove it.
This band starts just above the reach of the excitation tone, whose 9 harmonics end near 1.9\,kHz for this voice.
Above that frequency the generator has to synthesize the harmonics itself, and the small student gets their timing slightly wrong.

Changes to training did not remove the buzz.
Wider critics, direct waveform and phase losses, distilling the teacher's final layer, and larger decoders all left it audible.

We therefore correct the phase after synthesis instead, through a phase-locking filter.
The decoder already produces its own excitation tone, whose harmonics have steady phase.
In voiced frames and in the 2--8\,kHz band, the filter measures how far the output's phase drifts from this tone, smooths that drift over about a third of a second, and resynthesizes the audio with the smoothed phase and the original magnitude.
The idea is close to identity phase locking in the phase vocoder~\cite{laroche1999phasevocoder}.
The filter has no parameters and takes about 5\% of Paradee's run time.
By ear, it removes most of the buzz.
A slight buzz remains on some stressed syllables, where the student is 3--6\,dB quieter than the teacher between 2 and 8\,kHz, which no phase correction can fix.
UTMOS barely registers the change (4.39 to 4.41), and it did not track this artifact reliably elsewhere either, rating a larger decoder 0.5 lower than one that sounded the same by ear.
For this artifact we therefore relied on listening.

\section{Related work}
\label{sec:related}

\paragraph{Kokoro and its lineage.}
Kokoro-82M~\cite{hexgrad2025kokoro} uses the StyleTTS~2 architecture~\cite{li2023styletts2} without its diffusion and encoder parts.
StyleTTS~2 builds on StyleTTS~\cite{li2022styletts}, which introduced the style-conditioned duration, pitch and energy predictor that Paradee's text side imitates, and on PL-BERT~\cite{li2023plbert}, a phoneme-level ALBERT~\cite{lan2020albert}.
The decoder is an iSTFTNet~\cite{kaneko2022istftnet}, a HiFi-GAN~\cite{kong2020hifigan} whose last layers are replaced by a small inverse STFT, with Snake activations from BigVGAN~\cite{lee2023bigvgan} and a harmonic sine excitation from the neural source-filter family~\cite{wang2019nsf,wang2019hnnsf}.
HiFTNet~\cite{li2023hiftnet} combines the same two ideas as a standalone vocoder.
The hn-NSF model~\cite{wang2019hnnsf} learns a maximum voiced frequency above which the excitation is noise, which is directly relevant to our finding that the student's defect begins where Kokoro's fixed 9-harmonic excitation ends.

\paragraph{Distilling and compressing speech synthesis.}
Knowledge distillation~\cite{hinton2015distilling} entered TTS through FastSpeech~\cite{ren2019fastspeech}, which takes durations from an autoregressive teacher and trains on its outputs, and FastSpeech~2~\cite{ren2021fastspeech2}, which adds explicit pitch and energy predictors.
Paradee's text side is supervised the same way, but also on the teacher's phoneme features.
LightSpeech~\cite{luo2021lightspeech} profiled FastSpeech's components before compressing it 15$\times$, a precedent for our census.
Lai et al.~\cite{lai2022sparsity} studied how pruning TTS models and vocoders affects naturalness, intelligibility and prosody.
Vocoder distillation dates to Parallel WaveNet~\cite{oord2018parallelwavenet}.
More recently, Du et al.~\cite{du2025dllapnet} distilled a teacher's intermediate features into an amplitude-and-phase iSTFT vocoder, whereas distilling the teacher's iSTFT head did not help us (Section~\ref{sec:buzz}).
StyleTTS-ZS~\cite{li2024stylettszs} ``distills'' within the StyleTTS family, but it distills the diffusion sampler, not the model size.

The closest published work is Nix-TTS~\cite{chevi2022nixtts}, which distills VITS~\cite{kim2021vits} module by module into a 5.23M student that runs 3$\times$ faster on CPU.
sanoTTS~\cite{thapa2026sanotts} distills VITS voices from Piper into students under 2M parameters for microcontrollers, and notes that aggregate quality predictors missed a sibilant failure.
Both distill a VITS teacher, whose decoder is a HiFi-GAN without a harmonic source, and neither studies the excitation or phase effects we report.

\paragraph{Distilling Kokoro.}
We know of no paper that distills Kokoro or StyleTTS~2 into a small model.
Pipatanakul et al.~\cite{pipatanakul2026thai} train a full-size Kokoro as a single-voice Thai student on synthetic speech, so there Kokoro is the student and is not compressed.
The closest work is an unpublished model, Kokoro-7M-Distill~\cite{oddadmix2026kokoro7m}, released on Hugging Face in September 2026 while this work was under way.
It is a 7.48M single-voice English student of Kokoro-82M with the same broad shape as Paradee.
It uses teacher durations, a narrowed decoder, multi-resolution STFT and log-mel losses, multi-period and multi-resolution critics, and a WavLM feature-matching term~\cite{chen2022wavlm}.
Its training notes report UTMOS 4.14 and observe that without the WavLM term the output ``stays subtly robotic'', which may be the artifact we analyze.
The two efforts are independent.
Paradee differs in supervising the teacher's internal pitch, energy and phoneme features and in training the two halves separately.
It also adds an analysis of the teacher's parameters, compute and quantization sensitivity and the localization of the residual artifact.
On our 200 held-out sentences, with its matching voice pack, it scores UTMOS 4.18 against Paradee's 4.41 and runs about 1.4 times as fast (Table~\ref{tab:baselines}).

\paragraph{Small on-device TTS.}
Piper~\cite{hansen2023piper} ships per-voice VITS models and is the usual local baseline.
KittenTTS~\cite{kittenml2025kittentts} credits the StyleTTS~2 architecture.
Its smallest model has 14M parameters and eight voices, and it is not described as distilled.
SupertonicTTS~\cite{kim2025supertonictts} is an efficient 44M flow-matching system trained from scratch.
MB-iSTFT-VITS~\cite{kawamura2023mbistftvits} makes an end-to-end model light by using a multi-band iSTFT decoder.
For Kokoro itself, community int8 ONNX exports~\cite{onnxcommunity2025kokoroonnx} are about 86--92\,MB, against Paradee's 8.45\,MB.

\paragraph{Small iSTFT vocoders and phase.}
Vocos~\cite{siuzdak2023vocos} predicts a full-resolution spectrum and phase at frame rate and inverts it with one iSTFT, and iSTFTNet2~\cite{kaneko2023istftnet2} and WaveNeXt~\cite{okamoto2023wavenext} push the same direction.
APNet and APNet2~\cite{ai2023apnet,du2023apnet2} and FreeV~\cite{lv2024freev} predict amplitude and phase explicitly, and use anti-wrapping phase losses~\cite{ai2023antiwrapping} that handle the $\pm\pi$ wrap-around of phase.
Wavehax~\cite{yoneyama2024wavehax} argues that time-frequency vocoders lack a harmonic prior and adds one, which matches our diagnosis that the small decoder fails exactly where the harmonic prior stops.
The perceptual importance of phase in speech is well established~\cite{paliwal2011phase}, and loss of phase coherence across bins is the classical ``phasiness'' of the phase vocoder, which identity phase locking addresses~\cite{laroche1999phasevocoder}.
On the critic side, our multi-period and multi-resolution discriminators come from HiFi-GAN and UnivNet~\cite{kong2020hifigan,jang2021univnet} and see only waveforms and magnitudes.
EnCodec's~\cite{defossez2022encodec} and DAC's~\cite{kumar2023dac} discriminators look at complex spectra and split them into bands, and StyleTTS~2 adds a WavLM-based discriminator.
The multi-resolution STFT loss is from Parallel WaveGAN~\cite{yamamoto2020pwg}.

\paragraph{Evaluating naturalness automatically.}
UTMOS~\cite{saeki2022utmos} and its successor~\cite{baba2024utmosv2} predict MOS well on average, but recent work shows score-preserving degradations that UTMOS does not penalize~\cite{huang2026attackutmos} and that reference-free metrics do not reliably pick the clip listeners prefer among modern TTS systems~\cite{asonitis2026limits}.
Our results are a concrete case.
UTMOS rated one decoder 0.5 lower than another that sounded the same by ear, and it scores audibly buzzy output only 0.13 below the teacher.

\section{Limitations}

Paradee speaks only a single English voice, and we did not train it to speak any other voices or languages.
It can only be as good as its teacher and inherits the teacher's mispronunciations.
The quality evidence rests on an automatic predictor over 200 sentences and on informal listening by the author, rather than a formal listening test.
Besides the teacher, we compare against only three small open systems, each speaking a different voice.
Speed was measured on one machine only.
Paradee keeps a slight buzz on some voiced sounds, caused by phase and loudness errors in the 2--8\,kHz band.

\section{Conclusion}

An application that needs one voice does not need most of Kokoro.
Kokoro's architecture with much narrower layers, its two halves trained separately against the frozen teacher on one laptop, keeps most of its naturalness at a tenth of the parameters and a fifteenth of the compute.
The text side compresses easily.
The difficulty is concentrated in the waveform generator, and specifically in the phase of voiced speech just above the band that the excitation signal covers.
A filter that locks that phase to the model's own excitation, with no training, removes most of the resulting buzz.
We hope the teacher census, the quantization map and the buzz diagnosis save others the same search.
Code is available at \href{https://github.com/sahilmahendrakar/paradee}{github.com/sahilmahendrakar/paradee}, and the model files and audio samples at \href{https://huggingface.co/sahilmahendrakar/Paradee-8M-v1.0}{huggingface.co/sahilmahendrakar/Paradee-8M-v1.0}.

\bibliographystyle{plainnat}
\bibliography{refs}

\clearpage
\appendix
\section*{Appendix}
\addcontentsline{toc}{section}{Appendix}
\section{DTW log-mel distance}
\label{app:metric}

The distance is the mean absolute difference between RMS-normalized log-mel spectrograms after dynamic-time-warping alignment, so a rendition that is slightly faster or slower is not penalized for timing alone.
It is not gain-invariant without RMS normalization.
The ONNX export of the teacher is 1.44$\times$ louder than the PyTorch package, which at first showed up as a 0.69 gap.
Table~\ref{tab:calib} gives reference points on the same 5 sentences.
Two renditions at 0.39 could not be told apart by ear, and one at 0.94 sounded close to the teacher but slightly odd in places, so audible differences begin somewhere between about 0.4 and 0.9.

\begin{table}[h]
\centering
\small
\begin{tabular}{lrr}
\toprule
Reference point & DTW log-mel & Duration change \\
\midrule
Teacher against itself, second run (random excitation) & 0.09 & 0 \\
All weights fp16 & 0.11 & 0 \\
ONNX fp32 export against PyTorch & 0.21 & 0 \\
All weights int8 per channel, activations fp32 & 0.28 & 25 ms \\
Teacher at speed 0.97 (timing change only) & 0.36 & 130 ms \\
Teacher on CPU against Apple MPS & 0.39 & 0 \\
Prebuilt dynamic-int8 ONNX (activations quantized) & 0.58 & 65 ms \\
All weights int4 group-64 & 0.67 & 60 ms \\
A different voice (af\_bella), same text & 1.27 & 445 ms \\
\bottomrule
\end{tabular}
\caption{Reference points for the distance metric.}
\label{tab:calib}
\end{table}

\section{Quantization sensitivity of the teacher}
\label{app:quant}

We quantized the weights of one part of Kokoro at a time, keeping activations in fp32, and measured the change in the output with the distance of Appendix~\ref{app:metric} (Table~\ref{tab:quant}).
Sensitivity runs opposite to size.
The decoder's convolutional blocks, 41\% of the parameters, are unchanged at 8 bits and change little at 4 bits.
The text side changes more at 8 bits, because a nudged duration moves every later frame.
Quantizing all weights to 8 bits shifts durations by 25\,ms on average.
The prebuilt dynamic-int8 ONNX file, which also quantizes activations, shifts them by 65\,ms and doubles the distance.
All weights in fp16 are indistinguishable from fp32.

\begin{table}[h]
\centering
\small
\begin{tabular}{lrrrrr}
\toprule
Part & Params & 8-bit & 4-bit (groups) & 4-bit & Timing shift, 8-bit \\
\midrule
ALBERT & 6.3M & 0.263 & 0.495 & 0.555 & 10 ms \\
Text encoder & 5.6M & 0.153 & 0.266 & 0.319 & 5 ms \\
Prosody predictor & 16.2M & 0.312 & 0.379 & 0.402 & 25 ms \\
Decoder AdaIN & 33.6M & 0.090 & 0.126 & 0.179 & 0 \\
Generator & 19.7M & 0.103 & 0.354 & 0.537 & 0 \\
\midrule
All weights & 81.8M & 0.270 & 0.624 & & 25 ms \\
\bottomrule
\end{tabular}
\caption{Quantizing the weights of one part of Kokoro at a time, with the rest at full precision (8-bit per channel, 4-bit in groups of 64, 4-bit per channel). Each value is the DTW log-mel distance from the full-precision teacher on 5 sentences. Lower is better, and 0.09, the teacher's difference from its own second run, means no real change. The 0.4M projection after ALBERT is not quantized on its own. Snake activation parameters are kept in fp32, because quantizing them rounds small values to zero and produces NaNs.}
\label{tab:quant}
\end{table}

\section{Decoder sizes}
\label{app:decoders}

\begin{table}[h]
\centering
\small
\begin{tabular}{lrrrr}
\toprule
Preset & Widths and kernels & Params & Generator & GFLOP/s audio \\
\midrule
Teacher & 1024 / 512 / 3,7,11 & 53.3M & 19.7M & 51.9 \\
A (Paradee) & 256 / 128 / 3,7,11 & 3.85M & 1.19M & 3.36 \\
B & 256 / 128 / 3,7 & 3.49M & 0.83M & 2.28 \\
C & 192 / 96 / 3,7 & 2.16M & 0.48M & 1.31 \\
D & 128 / 64 / 3,7 & 1.15M & 0.23M & 0.61 \\
W2 (wider final stage) & 128 / 256 / 3,7 & 4.39M & 3.12M & 8.2 \\
L & 384 / 192 / 3,7,11 & 7.86M & 2.58M & 7.44 \\
\bottomrule
\end{tabular}
\caption{Student decoders, each a copy of Kokoro's decoder with scaled widths.}
\label{tab:decoders}
\end{table}

\end{document}